\documentclass[english]{article}
\usepackage[T1]{fontenc}
\usepackage[latin9]{inputenc}
\usepackage{geometry}
\usepackage{amstext}
\usepackage{esint}

\makeatletter

\usepackage{pdfpages}

\usepackage{fancyhdr, blindtext}

\newcommand{\changefont}{%

    \fontsize{9}{11}\selectfont}

{\changefont \slshape \rightmark}
{\changefont \slshape \leftmark}
\makeatother

\usepackage{babel}
\begin{document}

\section{Total force from the electromagnetic field of the source}

\subsection{Electric force}

The electric force on a dipolar particle of polarizability $\alpha_e$ is is \cite{ChaumetOL2001}
\begin{equation}
F_{i}^{e}\left(\mathbf{r}\right)=\frac{\varepsilon_{0}\varepsilon_{1}}{2}\Re\left\{ \left\langle \alpha_{e}E_{j}^{*}\left(\mathbf{r}\right)\partial_{i}E_{j}\left(\mathbf{r}\right)\right\rangle \right\} .
\end{equation}
The electric field from the source is \cite{Sipe1987green}
\begin{equation}
E_{i}^{inc}\left(\mathbf{r}\right)=\mu_{0}\mu_{2}\omega^{2}\int_{V}G_{ij}^{E,P}\left(\mathbf{r},\mathbf{r}',\omega\right)P_{j}\left(\mathbf{r}',\omega\right)d^{3}\mathbf{r}'.
\end{equation}
where $G_{ij}^{E,P}\left(\mathbf{r},\mathbf{r}',\omega\right)$ is
the electric Green function from the polarization currents
 and contains the transmission
Fresnel coefficients ($t_{s,p}$). $G_{ij}^{E,P}\left(\mathbf{r},\mathbf{r}',\omega\right)$
is expressed on using the angular spectrum of plane waves \cite{Sipe1987green,nietolibro}
\begin{eqnarray}
G_{ij}^{E,P}\left(\mathbf{r},\mathbf{r}',\omega\right) & = & \frac{i}{2}\int\frac{d^{2}\mathbf{K}}{\left(2\pi\right)^{2}}G_{ij}^{E,P}\left(\mathbf{K}\right)e^{i\mathbf{K}\left(\mathbf{R}-\mathbf{R}'\right)}e^{i\gamma_{1}z-i\gamma_{2}z'}.
\end{eqnarray}
where $G_{ij}^{E,P}\left(\mathbf{K}\right)=\frac{1}{\gamma_{2}}\left(\hat{s}_{i}t_{21}^{s}\hat{s}_{j}+\hat{p}_{1i}^{+}t_{21}^{p}\hat{p}_{2j}^{+}\right)$,
$\hat{\mathbf{s}}=\mathbf{\hat{K}}\times\hat{\mathbf{z}}$ , $\hat{\mathbf{p}}_{i}^{\pm}=-\left[\gamma_{i}\mathbf{\hat{K}}\mp K\hat{\mathbf{z}}\right]/(n_{i}k_{0})$
, $\gamma_{i}=\sqrt{\varepsilon_{i}\mu_{i}k_{0}^{2}-K^{2}}$ and $n_{i}=\sqrt{\varepsilon_{i}\mu_{i}}$
. $n_1$ ($n_2$) is the refractive index of the medium placed at $z<0$ ($z>0$). 

The self-correlation function $\left\langle E_{j}^{*}\left(\mathbf{r}\right)E_{j}\left(\mathbf{r}\right)\right\rangle $
at the position of the particle, i.e., at $\mathbf{r}=\mathbf{r}_{0}$
will be 
\begin{equation}
\left\langle E_{j}^{*}\left(\mathbf{r}_{0}\right)E_{j}\left(\mathbf{r}_{0}\right)\right\rangle =\mu_{0}^{2}\mu_{2}^{2}\omega^{4}\int_{V_{1},V_{2}}G_{ij}^{E,P*}\left(\mathbf{r},\mathbf{r}',\omega\right)G_{ij}^{E,P}\left(\mathbf{r},\mathbf{r}',\omega\right)W_{kl}^{(P)}\left(\mathbf{r}'_{1},\mathbf{r}'_{2},\omega\right)d^{3}\mathbf{r}'_{1}d^{3}\mathbf{r}'_{2},
\end{equation}
$W_{kl}^{(P)}\left(\mathbf{r}'_{1},\mathbf{r}'_{2},\omega\right)=\left\langle P_{k}^{*}\left(\mathbf{r}'_{1}\right)P_{l}\left(\mathbf{r}'_{2}\right)\right\rangle $
being the cross-spectral density tensor of the source polarization. Substituting
$W_{kl}^{(P)}\left(\mathbf{r}'_{1},\mathbf{r}'_{2},\omega\right)$
into Eq. (4) we have $\left\langle E_{j}^{*}\left(\mathbf{r}\right)E_{j}\left(\mathbf{r}\right)\right\rangle $
explicitly expressed in terms of the angular spectrum

\begin{eqnarray}
\left\langle E_{j}^{*}\left(\mathbf{r}_{0}\right)E_{j}\left(\mathbf{r}_{0}\right)\right\rangle  & = & \frac{\mu_{0}^{2}\mu_{2}^{2}\omega^{4}}{4}\int_{V_{1},V_{2}}\left[\int_{\mathbf{K}_{1},\mathbf{K}_{2}}\frac{d^{2}\mathbf{K}_{1}}{\left(2\pi\right)^{2}}\frac{d^{2}\mathbf{K}_{2}}{\left(2\pi\right)^{2}}G_{ij}^{E,P*}\left(\mathbf{r},\mathbf{r}',\omega\right)G_{ij}^{E,P}\left(\mathbf{r},\mathbf{r}',\omega\right)e^{-i\mathbf{K}_{1}\left(\mathbf{R}-\mathbf{R}_{1}'\right)}e^{i\mathbf{K}_{2}\left(\mathbf{R}-\mathbf{R}_{2}'\right)}\right.\nonumber \\
 & \times & \left.e^{-\left(i\gamma_{1,1}^{*}z-i\gamma_{2,1}^{*}z_{1}'\right)}e^{i\gamma_{1,2}z-i\gamma_{2,2}z_{2}'}\right]W_{kl}^{(P)}\left(\mathbf{r}'_{1},\mathbf{r}'_{2},\omega\right)d^{3}\mathbf{r}'_{1}d^{3}\mathbf{r}'_{2}.
\end{eqnarray}
We shall assume the correlation  with a Gaussian profile \cite{mandel1995optical}
\begin{equation}
W_{ij}^{(P)}\left(\mathbf{r}'_{1},\mathbf{r}'_{2},\omega\right)={\cal S}^{(P)}\left(\omega\right)\text{exp}\left(-(\left|\mathbf{r}'_{1}-\mathbf{r}'_{2}\right|)^{2}/2\sigma^{2}\right)\delta_{ij}/\left(2\pi\right)^{3/2}\sigma^{3},
\end{equation}
where ${\cal S}^{(P)}\left(\omega\right)$ is the normalized spectrum
of the source and $\sigma$ represents its correlation length of the source.
With a suitable
change of variables, the integration over the lateral space coordinates
gives a two-dimensional delta function $\delta^{(2)}\left[\mathbf{K}_{1}-\mathbf{K}_{2}\right]$ which
will simplify one of the integrations over $\mathbf{K}$. The rest
of integrals  lead to a more simplified equation: 
\begin{eqnarray}
\left\langle E_{j}^{*}\left(\mathbf{r}_{0}\right)E_{j}\left(\mathbf{r}_{0}\right)\right\rangle  & = & \frac{\mu_{0}^{2}\mu_{2}^{2}\omega^{4}}{4}\frac{1}{\left(2\pi\right)^{6/5}}{\cal S}^{(P)}\left(\omega\right)\nonumber \\
 & \times & \int_{\mathbf{K}}\frac{1}{\left|\gamma_{2}\right|^{2}}e^{-\frac{\left(\mathbf{K}\sigma\right)^{2}}{2}}\left[\left|t_{21}^{s}\right|^{2}+\left|t_{21}^{p}\right|^{2}\frac{1}{n_{1}^{2}k_{0}^{2}}\left(\left|\gamma_{1}\right|^{2}+K^{2}\right)\frac{1}{n_{2}^{2}k_{0}^{2}}\left(\left|\gamma_{2}\right|^{2}+K^{2}\right)\right]\nonumber \\
 &  & e^{-2z_{0}\Im\gamma_{1}}\frac{1}{2\Im\gamma_{2}}e^{-\frac{1}{2}\sigma^{2}\Re\gamma_{2}^{2}}d^{2}\mathbf{K}.
\end{eqnarray}

Next,  we will divide the force into conservative and non-conservative components.

\subsubsection{Conservative force.}

The conservative electric force  $F_{i}^{e,cons}=\text{Re}\alpha\partial_{i}\left\langle E_{j}^{*}\left(\mathbf{r}\right)E_{j}\left(\mathbf{r}\right)\right\rangle /4$ due to the electric field
is  given by 
\begin{eqnarray}
F_{z}^{e-cons} & = & -\frac{\varepsilon_{0}\varepsilon_{1}}{4}\Re\left\{ \alpha_{e}\right\} \frac{\mu_{0}^{2}\mu_{2}^{2}\omega^{4}}{4}\frac{1}{\left(2\pi\right)^{1/5}}{\cal S}^{(P)}\left(\omega\right)\nonumber \\
 & \times & \int_{K=k_{0}}^{K=+\infty}2\sqrt{K^{2}-k_{0}^{2}}\frac{1}{\left|\gamma_{2}\right|^{2}}e^{-\frac{\left(\mathbf{K}\sigma\right)^{2}}{2}}\left[\left|t_{21}^{s}\right|^{2}+\left|t_{21}^{p}\right|^{2}\frac{1}{n_{1}^{2}k_{0}^{2}}\left(\left|\gamma_{1}\right|^{2}+K^{2}\right)\frac{1}{n_{2}^{2}k_{0}^{2}}\left(\left|\gamma_{2}\right|^{2}+K^{2}\right)\right]\nonumber \\
 & \times & e^{-2z_{0}\Im\gamma_{1}}\frac{1}{2\Im\gamma_{2}}e^{-\frac{1}{2}\sigma^{2}\Re\gamma_{2}^{2}}KdK,
\end{eqnarray}
where only the third Cartesian component  and  the
contribution is solely due to the evanescent modes. This integration is 
numerically resolved .

\subsubsection{Non conservative force.}

The non-conservative force  $F_{i}^{e,nc}=\text{Im}\alpha\text{Im}\left\langle E_{j}^{*}\left(\mathbf{r}\right)\partial_{i}E_{j}\left(\mathbf{r}\right)\right\rangle /2$ due to the electric field
, is determined in a similar way, now calculating $\left\langle E_{j}^{*}\left(\mathbf{r}\right)\partial_{i}E_{j}\left(\mathbf{r}\right)\right\rangle $
instead of $\left\langle E_{j}^{*}\left(\mathbf{r}\right)\partial_{i}E_{j}\left(\mathbf{r}\right)\right\rangle $.
Hence, 

\begin{eqnarray}
F_{i}^{e-nc} & = & \frac{\varepsilon_{0}\varepsilon_{1}}{2}\text{Im}\alpha_{e}\frac{\mu_{0}^{2}\mu_{2}^{2}\omega^{4}}{4}\frac{1}{\left(2\pi\right)^{1/5}}{\cal S}^{(P)}\left(\omega\right)\nonumber \\
 & \times & \int_{K=0}^{K=k_{0}}\sqrt{k_{0}^{2}-K^{2}}\frac{1}{\left|\gamma_{2}\right|^{2}}e^{-\frac{\left(\mathbf{K}\sigma\right)^{2}}{2}}\left[\left|t_{21}^{s}\right|^{2}+\left|t_{21}^{p}\right|^{2}\frac{1}{n_{1}^{2}k_{0}^{2}}\left(\left|\gamma_{1}\right|^{2}+K^{2}\right)\frac{1}{n_{2}^{2}k_{0}^{2}}\left(\left|\gamma_{2}\right|^{2}+K^{2}\right)\right]\nonumber \\
 & \times & \frac{1}{2\Im\gamma_{2}}e^{-\frac{1}{2}\sigma^{2}\Re\gamma_{2}^{2}}KdK,
\end{eqnarray}
where only the homogeneous waves give a non-zero value which is constant
for any value of $\mathbf{r}$. This integration is resolved numerically.

\subsection{Magnetic force}

The magnetic force for a magnetodielectric particle is \cite{nieto2010optical}

\begin{equation}
F_{i}^{m}\left(\mathbf{r}\right)=\frac{\mu_{0}\mu_{1}}{2}\Re\left\{ \left\langle \alpha_{m}H_{j}^{*}\left(\mathbf{r}\right)\partial_{i}H_{j}\left(\mathbf{r}\right)\right\rangle \right\} .
\end{equation}
The magnetic field emitted by the source is
Maxwell's equations, 
\begin{equation}
H_{j}^{inc}\left(\mathbf{r}\right)=-i\omega\int_{V}G_{jk}^{H,P}\left(\mathbf{r},\mathbf{r}',\omega\right)P_{k}\left(\mathbf{r}',\omega\right)d^{3}\mathbf{r}',
\end{equation}
where the magnetic Green's function is 
\begin{eqnarray}
G_{jk}^{H,P}\left(\mathbf{r},\mathbf{r}',\omega\right) & = & \frac{k_{0}n_{2}}{2}\int\frac{d^{2}\mathbf{K}}{\left(2\pi\right)^{2}}G_{km}^{H,P}\left(\mathbf{K}\right)e^{i\mathbf{K}\left(\mathbf{R}-\mathbf{R}'\right)}e^{i\gamma_{1}z-i\gamma_{2}z'},
\end{eqnarray}
and $G_{kl}^{H}\left(\mathbf{K}\right)=\frac{1}{\gamma_{2}}\left(\hat{p}_{1k}^{+}t_{21}^{s}\hat{s}_{l}-\hat{s}_{k}t_{21}^{p}\hat{p}_{2l}^{+}\right)$.
The rest of the calculation is similar to the one described in  Section
1.1.

\subsection{Interaction force}

The  force from the interference between the electric and magnetic dipoles is \cite{nieto2010optical}

\begin{equation}
F_{i}^{e-m}\left(\mathbf{r}\right)=-\varepsilon_{0}\varepsilon_{1}\frac{Zk_{0}^{4}}{12\pi}\Re\left\{ (\alpha_{e}^{*}\alpha_{m})\left\langle \mathbf{E}^{*}\times\mathbf{H}\right\rangle _{i}\right\} ,
\end{equation}
 where $Z=\sqrt{\mu_{0}\mu_{1}/(\varepsilon_{0}\varepsilon_{1})}$.
Once we have defined the electric and magnetic fields emitted by
the source [cf. Eqs. (2-3) and (11-12)], one can calculate the cross
product $\left\langle \mathbf{E}^{*}\times\mathbf{H}\right\rangle $, 

\newpage

\section{Total force from the electromagnetic field emitted by the particle induced electric and magnetic dipoles}

\subsection{Electric force}

The field emitted by the electric (magnetic) dipole $p$ ($m$), after reflections on the source surface $z=0$ is 
\begin{eqnarray}
E_{i}^{p}\left(\mathbf{r}\right) & = & \mu_{0}\mu_{2}\omega^{2}\int_{V}G_{ij}^{E,p}\left(\mathbf{r},\mathbf{r}',\omega\right)p_{j}\left(\mathbf{r}',\omega\right)\delta\left(\mathbf{r}'-\mathbf{r}_{0}\right)d^{3}\mathbf{r}',\nonumber \\
 & = & \mu_{0}\mu_{2}\omega^{2}G_{ij}^{E,p}\left(\mathbf{r},\mathbf{r}_{0},\omega\right)p_{j}\left(\mathbf{r}_{0},\omega\right),
\end{eqnarray}
\begin{equation}
E_{i}^{m}\left(\mathbf{r}\right)=\frac{Z_{0}i\omega}{c}G_{ij}^{H,m\leftrightarrow}\left(\mathbf{r},\mathbf{r}',\omega\right)m_{j}\left(\omega\right),
\end{equation}
where the electric Green's function contains the properties of
the source through the reflection Fresnel coefficients ($r_{s,p}$)
\begin{eqnarray}
G_{ij}^{E,p}\left(\mathbf{r},\mathbf{r}',\omega\right) & = & \frac{i}{2}\int\frac{d^{2}\mathbf{K}}{\left(2\pi\right)^{2}}G_{ij}^{E,p}\left(\mathbf{K}\right)e^{i\mathbf{K}\left(\mathbf{R}-\mathbf{R}'\right)}e^{i\gamma_{1}\left(z+z'\right)},\\
G_{ij}^{H,m\leftrightarrow}\left(\mathbf{r},\mathbf{r}',\omega\right) & = & \frac{k_{0}n_{2}}{2}\int\frac{d^{2}\mathbf{K}}{\left(2\pi\right)^{2}}G_{ij}^{H,m\leftrightarrow}\left(\mathbf{K}\right)e^{i\mathbf{K}\left(\mathbf{R}-\mathbf{R}'\right)}e^{i\gamma_{1}\left(z+z'\right)}.
\end{eqnarray}
 $G_{ij}^{E,p}\left(\mathbf{K}\right)=\frac{1}{\gamma_{1}}\left(\hat{s}_{i}r_{12}^{s}\hat{s}_{j}+\hat{p}_{1i}^{+}r_{12}^{p}\hat{p}_{1j}^{-},\right)$
and $G_{ij}^{H,m\leftrightarrow}\left(\mathbf{K}\right)=\frac{1}{\gamma_{1}}\left(\hat{p}_{1i}^{+}r_{12}^{p}\hat{s}_{j}-\hat{s}_{i}r_{12}^{s}\hat{p}_{1j}^{-},\right)$.
The superscript $^{\leftrightarrow}$ denotes that the electric field
generated by the magnetic dipole has the same Green's function as
the magnetic field radiated by the magnetic dipole with the interchange $r_{s}\leftrightarrow r_{p}$.
Notice also that in the Green function described above there is
no  free-space term. This is due to the multiple scattering of the dipole field with the source surface,
(see for example \cite{novotny2006principles,sipe1987quantum})
.

The correlation function $\left\langle E_{i}^{*}\left(\mathbf{r}\right)E_{i}\left(\mathbf{r}\right)\right\rangle $
at the position of the particle is obtained on considering that the cross-correlation between the electric and magnetic dipoles is zero, i.e., $\left\langle p_{i}^{*}m_{j}\right\rangle =0$,
therefore
\begin{equation}
\left\langle E_{i}^{*}\left(\mathbf{r}_{0}\right)E_{i}\left(\mathbf{r}_{0}\right)\right\rangle =\left\langle \left(E_{i}^{p*}\left(\mathbf{r}_{0}\right)+E_{i}^{m*}\left(\mathbf{r}_{0}\right)\right)\left(E_{i}^{p}\left(\mathbf{r}_{0}\right)+E_{i}^{m}\left(\mathbf{r}_{0}\right)\right)\right\rangle =\left\langle E_{i}^{p*}\left(\mathbf{r}_{0}\right)E_{i}^{p}\left(\mathbf{r}_{0}\right)\right\rangle +\left\langle E_{i}^{m*}\left(\mathbf{r}_{0}\right)E_{i}^{m}\left(\mathbf{r}_{0}\right)\right\rangle ,
\end{equation}
 each ensemble average being

\begin{eqnarray}
\left\langle E_{j}^{p*}\left(\mathbf{r}_{0}\right)E_{j}^{p}\left(\mathbf{r}_{0}\right)\right\rangle  & = & \mu_{0}^{2}\mu_{1}^{2}\omega^{4}G_{jk}^{E,p*}\left(\mathbf{r}_{0},\mathbf{r}_{0},\omega\right)G_{jl}^{E,p}\left(\mathbf{r},\mathbf{r}',\omega\right)\left\langle p_{k}^{*}\left(\mathbf{r}_{0},\omega\right)p_{l}\left(\mathbf{r}_{0},\omega\right)\right\rangle ,\\
\left\langle E_{j}^{m*}\left(\mathbf{r}_{0}\right)E_{j}^{m}\left(\mathbf{\mathbf{r}_{0}}\right)\right\rangle  & = & \left(\frac{Z_{0}\omega}{c}\right)^{2}G_{jk}^{H,m\leftrightarrow*}\left(\mathbf{r}_{0},\mathbf{r}_{0},\omega\right)G_{jl}^{H,m\leftrightarrow}\left(\mathbf{r},\mathbf{r}',\omega\right)\left\langle m_{k}^{*}\left(\mathbf{r}_{0},\omega\right)m_{l}\left(\mathbf{r}_{0},\omega\right)\right\rangle ,
\end{eqnarray}
and

\begin{eqnarray}
p_{k}\left(\mathbf{r},\omega\right) & = & \varepsilon_{0}\varepsilon_{1}\alpha_{e}\left(\omega\right)E_{k}^{inc}\left(\mathbf{\mathbf{r}_{0}},\mathbf{r}_{1},\omega\right),\\
p_{k}\left(\mathbf{r},\omega\right) & = & \alpha_{m}\left(\omega\right)H_{k}^{inc}\left(\mathbf{\mathbf{r}_{0}},\mathbf{r}_{1},\omega\right).
\end{eqnarray}
The correlation tensor $\left\langle E_{k}^{inc*}\left(\mathbf{r}_{0},\mathbf{r}_{1},\omega\right)E_{l}^{inc}\left(\mathbf{r}_{0},\mathbf{r}_{1},\omega\right)\right\rangle $
has been calculated in the Section 1.1, where $\mathbf{r}_{1}$ is
a point of the half-space ocupied by source. After a somewhat protracted calculation one obtains that at the particle position one has that   $\left\langle E_{j}^{m*}\left(\mathbf{r}_{0}\right)E_{j}^{m}\left(\mathbf{r}_{0}\right)\right\rangle =0$,
thus the only term remaining different from zero is $\left\langle E_{j}^{p*}\left(\mathbf{r}_{0}\right)E_{j}^{p}\left(\mathbf{r}_{0}\right)\right\rangle $:

\begin{eqnarray}
\left\langle E_{j}^{p*}\left(\mathbf{r}_{0}\right)E_{j}^{p}\left(\mathbf{r}_{0}\right)\right\rangle  & = & \mu_{0}^{2}\mu_{1}^{2}\omega^{4}\left|\varepsilon_{0}\right|^{2}\left|\varepsilon_{1}\right|^{2}\left|\alpha_{e}\right|^{2}G_{jk}^{E,p*}\left(\mathbf{r}_{0},\mathbf{r}_{0},\omega\right)G_{jl}^{E,p}\left(\mathbf{r}_{0},\mathbf{r}_{0},\omega\right)\left\langle E_{k}^{inc*}\left(\mathbf{r}_{0},\mathbf{r}_{1}',\omega\right)E_{l}^{inc}\left(\mathbf{r}_{0},\mathbf{r}_{1}',\omega\right)\right\rangle \nonumber \\
 & = & \mu_{0}^{2}\mu_{1}^{2}\omega^{4}\left|\varepsilon_{0}\right|^{2}\left|\varepsilon_{1}\right|^{2}\left|\alpha_{e}\right|^{2}\frac{\mu_{0}^{2}\omega^{2}}{4}\frac{1}{\left(2\pi\right)^{1/5}}{\cal S}^{(P)}\left(\omega\right)\nonumber \\
 & \times & \left|\frac{i}{2}\int\frac{2\pi KdK}{\left(2\pi\right)^{2}}\frac{1}{\gamma_{1}}\left(r_{12}^{s}+\frac{r_{12}^{p}}{\left(n_{1}k_{0}\right)^{2}}\left(\gamma_{1}^{2}-K^{2}\right)\right)e^{i\gamma_{1}\left(z_{0}+z_{1}'\right)}\right|^{2}\nonumber \\
 & \times & \int_{K}\frac{2\pi}{\left|\gamma_{2}\right|^{2}}e^{-\frac{\left(\mathbf{K}\sigma\right)^{2}}{2}}\left(\left|t_{21}^{s}\right|^{2}+\left|t_{21}^{p}\right|^{2}\frac{1}{n_{1}^{2}n_{2}^{2}k_{0}^{4}}\left(\left|\gamma_{1}\right|^{2}+K^{2}\right)\left(\left|\gamma_{2}\right|^{2}+K^{2}\right)\right)e^{-2z_{0}\Im\gamma_{1}}\frac{1}{2\Im\gamma_{2}}e^{-\frac{1}{2}\sigma^{2}\Re\gamma_{2}^{2}}KdK.\nonumber \\
\end{eqnarray}
After completing the whole procedure, we derive the 
conservative plus the non-conservative parts of the total electric force.

\subsection{Magnetic force}

The magnetic field emerged from either the electric and the magnetic dipoles is calculated by using the following substitutions \cite{jackson1998classical}:
\begin{eqnarray}
E_{i}^{m}\left(\mathbf{r}\right) & = & -\frac{Z_{0}}{c}H_{i}^{p}\left(\mathbf{r}\right),\\
H_{i}^{m}\left(\mathbf{r}\right) & = & \frac{1}{Z_{0}c}E_{i}^{p}\left(\mathbf{r}\right),\\
\mathbf{p} & \rightarrow & \mathbf{m},\\
r^{s} & \leftrightarrow & r^{p,}
\end{eqnarray}
The procedure to obtain the force follows the same steps as described in Section 2.1

\subsection{Interaction force}

Once whe have characterized the electric and magnetic fields emerging from
theelectric and magnetic dipoles [cf.  Eqs. (14-17)], one  calculates the cross product
$\left\langle \mathbf{E}^{*}\times\mathbf{H}\right\rangle $ in a
way similarto that of see section 2.1.

\bibliographystyle{unsrt}

\end{document}